\documentclass[showkeys,superscriptaddress,showpacs,preprintnumbers,aps,twocolumn,prl]{revtex4}
\usepackage[latin1]{inputenc}
\usepackage{epsfig}
\usepackage{amsfonts}
\usepackage{amssymb}
\usepackage{amsmath}
\usepackage{color}
\newcommand{\ITP}{Institut f{\"u}r Theoretische Physik, 
  Technische Universit{\"a}t Berlin,
  Hardenbergstra{\ss}e 36, 10623 Berlin, Germany}
\newcommand{\FU}{Institut f{\"u}r Mathematik I, FU Berlin, Arnimallee 2-6, D-14195 Berlin, Germany}

\renewcommand{\Re}{\mathrm{Re}}

\begin{document}
\preprint{Phys. Rev. Lett. (2007), in print}
\date{\today}

\title{Refuting the odd number limitation of time-delayed feedback control}
\author{B. Fiedler} \affiliation{\FU} \author{V. Flunkert}
\affiliation{\ITP} \author{M. Georgi} \affiliation{\FU} \author{P.
  H{\"o}vel} \affiliation{\ITP} \author{E. Sch\"oll}
\affiliation{\ITP}

\begin{abstract}
  {We refute an often invoked theorem which claims that a periodic
    orbit with an odd number of real Floquet multipliers greater than
    unity can never be stabilized by time-delayed feedback control in
    the form proposed by Pyragas.  Using a generic normal form, we
    demonstrate that the unstable periodic orbit generated by a
    subcritical Hopf bifurcation, which has a single real unstable
    Floquet multiplier, can in fact be stabilized.  We derive explicit
    analytical conditions for the control matrix in terms of the
    amplitude and the phase of the feedback control gain, and present
    a numerical example.  Our results are of relevance for a wide
    range of systems in physics, chemistry, technology, and life
    sciences, where subcritical Hopf bifurcations occur.}

\end{abstract}

 \maketitle

The stabilization of unstable and chaotic systems is a central issue
in applied nonlinear science \cite{SCH99c,BOC00,GAU03a}.  Starting
with the work of Ott, Grebogi and Yorke \cite{OTT90}, a variety of
methods have been developed in order to stabilize unstable periodic
orbits (UPOs) embedded in a chaotic attractor by employing tiny
control forces.  A particularly simple and efficient scheme is
time-delayed feedback as suggested by Pyragas \cite{PYR92}.
It is an attempt to stabilize periodic orbits of minimal period $T$ by
a feedback control which involves a time delay $\tau=nT$, for suitable
positive integer $n$. A linear feedback example is
\begin{eqnarray}
        \dot z(t) =f(\lambda, z(t)) + B [z(t-\tau)-z(t)] \label{eq:ex}
\end{eqnarray}
where $\dot z(t) =f(\lambda, z(t))$ describes a d-dimensional
nonlinear dynamical system with bifurcation parameter $\lambda$ and an
unstable orbit of period $T$. $B$ is a suitably chosen constant
feedback control matrix. Typical choices are multiples of the identity
or of rotations, or matrices of low rank. More general nonlinear
feedbacks are conceivable, of course.  The main point, however, is
that the Pyragas choice $\tau_P=nT$ of the delay time eliminates the
feedback term in case of successful stabilization and thus recovers
the original $T$-periodic solution $z(t)$.  In this sense the method
is noninvasive.
Although time delayed feedback control has been widely used with great
success in real world problems in physics, chemistry, biology, and
medicine, e.g.
\cite{PYR93,BIE94,PIE96,HAL97,SUK97,LUE01,PAR99,KRO00a,FUK02,LOE04,ROS04a,POP05,SCH06a},
severe limitations are imposed by the common belief that certain
orbits cannot be stabilized for any strength of the control force. In
fact, it has been contended that periodic orbits with an odd number of
real Floquet multipliers greater than unity cannot be stabilized by
the Pyragas method \cite{JUS97,NAK97,NAK98,HAR01,PYR04a,PYR06}, even
if the simple scheme (\ref{eq:ex}) is extended by multiple delays in
form of an infinite series \cite{SOC94}.  To circumvent this
restriction other, more complicated, control schemes, like an
oscillating feedback \cite{SCH97p},
or the introduction of an additional, unstable degree of freedom
\cite{PYR01,PYR06}, have been proposed.  In this letter, we claim, and
show by example, that the general limitation for orbits with an odd
number of real unstable Floquet multipliers greater than unity does
not hold, but that stabilization may be possible for suitable choices
of $B$. We illustrate this with an example which consists of an
unstable periodic orbit generated by a subcritical Hopf bifurcation,
refuting the theorem in \cite{NAK97}.

Consider the normal form of a subcritical Hopf bifurcation, extended by a
time delayed feedback term
\begin{eqnarray}
        \dot z(t) = \left[\lambda + i +(1+ i \gamma)|z(t)|^2\right] z(t) + b [z(t-\tau)-z(t)] \label{eq:sub}
\end{eqnarray}
with $z \in {\mathbb C}$ and real parameters $\lambda$ and $\gamma$.
Here the Hopf frequency is normalized to unity.  The feedback matrix
$B$ is represented by multiplication with a complex number
$b=b_R+ib_I=b_0 e^{i\beta}$ with real $b_R,b_I,\beta$, and positive
$b_0$. Note that the nonlinearity $f(\lambda,z(t))= \left[\lambda + i
  +(1+ i \gamma)|z(t)|^2\right] z(t)$ commutes with complex rotations.
Hence the Hopf bifurcations from the trivial solution $z \equiv0$ at
simple imaginary eigenvalue $\eta=i\omega \neq 0$ produce rotating
wave solutions $z(t) = z(0)\exp\left(i \frac{2 \pi}{T}t\right)$ with
period $T$ even in the nonlinear case and with delay terms. This
follows from uniqueness of the emanating Hopf branches.  

Transforming Eq.~(\ref{eq:sub}) to amplitude and phase variables
$r,\theta$ using $z(t)=r(t) e^{i\theta(t)}$, we obtain at $b=0$
\begin{eqnarray}
        \dot r(t) &=& \left(\lambda + r^2\right) r \\
        \dot \theta(t) &=& 1 + \gamma r^2.
\end{eqnarray}
An unstable periodic orbit (UPO) with $r^2=-\lambda$ and period $T=2
\pi/(1-\gamma \lambda)$ exists for $\lambda<0$.  At $\lambda=0$ a
subcritical Hopf bifurcation occurs. The Pyragas control method
chooses delays as $\tau_P=nT$. This defines the local {\em Pyragas
  curve} in the $(\lambda,\tau)$-plane for any $n \in {\mathbb N}$
\begin{eqnarray}
        \tau_P(\lambda)=\frac{2\pi n}{1-\gamma \lambda} = 2\pi n (1+\gamma \lambda+\dots)\label{eq:tau_Pyragas}
\end{eqnarray}
which emanates from the Hopf bifurcation point $\lambda=0$.  Under
further nondegeneracy conditions, the Hopf point $\lambda=0$,
$\tau=nT$ ($n \in {\mathbb N}_0$) continues to a Hopf bifurcation
curve $\tau_H(\lambda)$ for $\lambda <0$. We determine this {\em Hopf
  curve} next. It is characterized by purely imaginary eigenvalues
$\eta = i \omega$ of the transcendental characteristic equation
\begin{eqnarray}
        \eta = \lambda + i + b\left( e^{-\eta \tau}-1 \right) \label{eq:char}
\end{eqnarray} 
which results from the linearization
at the steady state $z=0$ of the delayed system (\ref{eq:sub}).

\begin{figure}[ht]
  \centering \epsfig{file=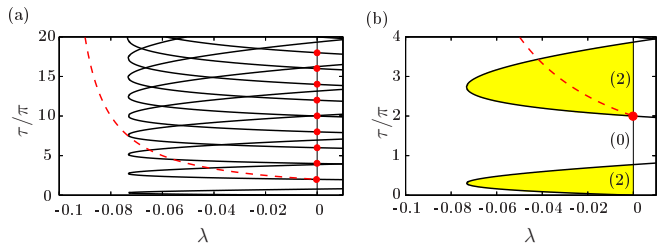,width=\columnwidth}
  \caption{(Color online) Pyragas (red dashed) and Hopf (black solid) curves in the
    $(\lambda,\tau)$-plane: (a) Hopf bifurcation curves $n=0,..., 10$,
    (b) Hopf bifurcation curves $n=0,1$ in an enlarged scale.  Yellow
    shading marks the domains of unstable $z=0$ and numbers in
    parentheses denote the dimension of the unstable manifold of $z=0$
    ($\gamma = -10$, $ b_0=0.3$ and $\beta=\pi/4$).
  }\label{fig:lambda_tau}
\end{figure}

\begin{figure}[ht]
  \centering \epsfig{file=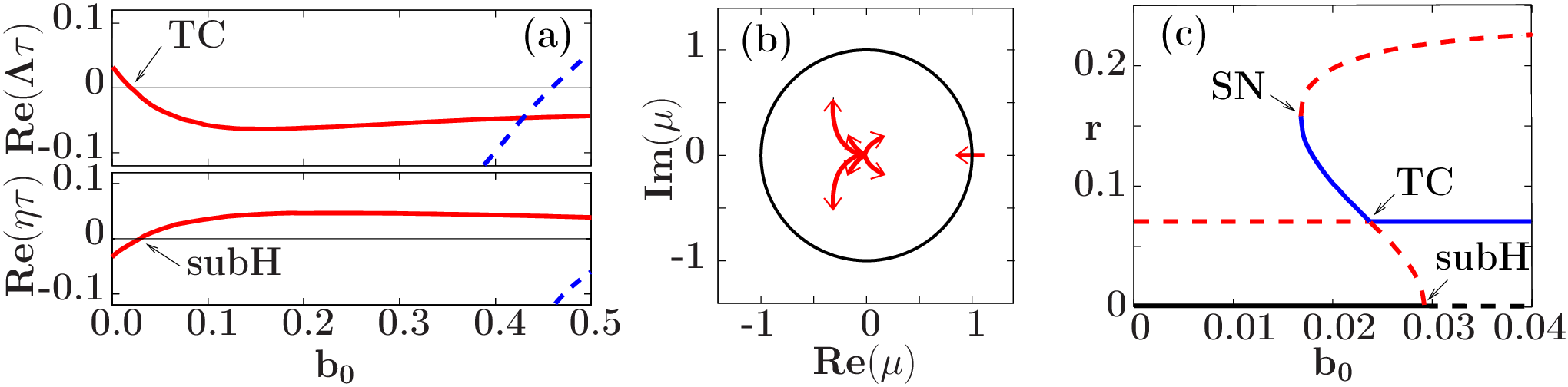,width=\columnwidth}
  \caption{(Color online) (a) top: Real part of Floquet exponents 
    $\Lambda$ of the periodic orbit vs. feedback amplitude $b_0$.
    bottom: Real part of eigenvalue $\eta$ of steady state vs.
    feedback amplitude $b_0$.  
    (b): Floquet multipliers $\mu= \exp(\Lambda \tau)$ (red) in the complex
    plane with the feedback amplitude $b_0 \in [0, 0.3]$ as a
    parameter.  (c): radii of periodic orbits. Solid
    (dashed) lines correspond to stable (unstable) orbits.  
    ($\lambda= -0.005$, $\gamma = -10$, $\tau=\frac{2
    \pi}{1-\gamma \lambda}$, $\beta=\pi/4$).  }\label{fig:spectrum}
\end{figure}

Separating Eq.~(\ref{eq:char}) into real and imaginary parts
\begin{eqnarray}
0&=&\lambda+b_0[\cos(\beta-\omega\tau)-\cos\beta]\\
\omega-1&=&b_0[\sin(\beta-\omega\tau)-\sin\beta]
\end{eqnarray} 
and using trigonometric identities to eliminate $\omega(\lambda)$
yields an explicit expression for the multivalued Hopf curve
$\tau_H(\lambda)$ for given control amplitude $b_0$ and phase $\beta$:
\begin{eqnarray}
        \tau_H = \frac{ \pm \arccos\left(\frac{b_0 \cos\beta-\lambda}{b_0}\right)+\beta+ 2 \pi n}{1- b_0\sin\beta
\pm \sqrt{\lambda (2 b_0 \cos\beta-\lambda)+b_0^2 \sin^2 \beta }}. \label{eq:tau_Hopf}
\end{eqnarray} 
Note that $\tau_H$ is not defined in case of $\beta=0$ and
$\lambda<0$. Thus complex $b$ is a necessary condition for the
existence of the Hopf curve in the subcritical regime $\lambda<0$.
Fig.~\ref{fig:lambda_tau} displays the family of Hopf curves, $n \in
{\mathbb N}_0$, Eq. (\ref{eq:tau_Hopf}), and the Pyragas curve $n=1$,
Eq. (\ref{eq:tau_Pyragas}), in the ($\lambda,\tau$) plane. In
Fig.~\ref{fig:lambda_tau}(b) the domains of instability of the trivial
steady state $z=0$, bounded by the Hopf curves, are marked by light
grey shading (yellow online). The dimensions of the unstable manifold
of $z=0$ are given in parentheses along the $\tau$-axis in
Fig.~\ref{fig:lambda_tau}(b). By construction, the period of the
bifurcating periodic orbits becomes equal to $\tau_P=nT$ along the
Pyragas curve, since the time-delayed feedback term vanishes. 
Standard exchange of stability results \cite{DIE95}, which hold verbatim for delay
equations, then assert that the bifurcating branch of periodic
solutions locally inherits linear asymptotic (in)stability from the trivial
steady state, i.e., it consists of stable periodic orbits on 
the Pyragas curve $\tau_P(\lambda)$ inside the yellow domains 
for small $|\lambda|$. Note that an unstable trivial steady state 
is not a sufficient condition for stabilization of the subcritical orbit, 
but other (e.g., global) bifurcations at $\lambda<0$ must be considered as well. 
More precisely, for small $|\lambda|$ 
the unstable periodic orbits possess a single Floquet multiplier $\mu= \exp(\Lambda
\tau) \in (1,\infty)$, near unity, which is simple.  All other
nontrivial Floquet multipliers lie strictly inside the complex unit
circle. In particular, the (strong) unstable dimension of these
periodic orbits is odd, here 1, and their unstable manifold is
two-dimensional. This is shown in Fig.~\ref{fig:spectrum}, which
depicts solutions $\Lambda$ of the characteristic equation of the
periodic solution on the Pyragas curve.
Panel (a) (top) shows the dependence of the real part of the critical
Floquet exponent $\Lambda$ on the amplitude of the feedback gain
$b_0$.  The largest real part is positive for $b_0=0$. Thus the
periodic orbit is unstable. As the amplitude of the feedback gain
increases, the largest real part of the eigenvalue becomes smaller and
eventually changes sign. Hence the periodic orbit is stabilized.  Note
that an infinite number of Floquet exponents are created by the
control scheme; their real parts tend to $-\infty$ in the limit $b_0
\to 0$, and some of them may cross over to positive real parts for
larger $b_0$ (blue curve), terminating the stability of the periodic
orbit.  
Panel (b) of Fig.~\ref{fig:spectrum} shows
the behavior of the Floquet multipliers $\mu=\exp(\Lambda \tau)$ in
the complex plane with the increasing amplitude of the feedback gain
$b_0$ as a parameter (marked by arrows).  There is an isolated real
multiplier crossing the unit circle at $\mu=1$, in contrast to the
result stated in \cite{NAK97}. This is caused by a transcritical
bifurcation 
(TC) in which the subcritical Pyragas orbit (whose radius is given
by $r=(-\lambda)^{1/2}$ independently of the control amplitude $b_0$) 
collides with a delay-induced periodic orbit, 
as shown in Fig.~\ref{fig:spectrum}(c). This delay-induced orbit
is generated at a finite value of the control amplitude $b_0$ (SN) by a 
saddle-node bifurcation
(collision with another unstable delay-induced periodic orbit).
At TC, the subcritical orbit and the delay-induced orbit exchange stability. 
The latter vanishes at a subcritical Hopf (subH) bifurcation at which 
the trivial steady state becomes unstable. 
Except at TC, the delay-induced orbit has a period $T\ne \tau$.
Note that for small $b_0$ the subcritical orbit is unstable, while $z=0$ 
is stable, but the respective exchanges of stability occur at slightly different
values of $b_0$, corresponding to TC and subH. This is
also corroborated by Fig.~\ref{fig:spectrum}(a) (bottom), which
displays the largest real part of the eigenvalues $\eta$ of the
steady state $z=0$.  
The possible existence of such delay-induced periodic orbits with 
$T \neq \tau$, 
which results in a Floquet multiplier $\mu=1$ of multiplicity two at TC, 
was overlooked in \cite{NAK97}.

Next we analyse the conditions under which stabilization of the
subcritical periodic orbit is possible. From
Fig.~\ref{fig:lambda_tau}(b) it is evident that the Pyragas curve must
lie inside the yellow region, i.e., the Pyragas and Hopf curves
emanating from the point $(\lambda, \tau)=(0, 2\pi)$ must locally
satisfy the inequality $\tau_H(\lambda)<\tau_P(\lambda)$ for
$\lambda<0$.  More generally, let us investigate the eigenvalue
crossings of the Hopf eigenvalues $\eta=i\omega$ along the $\tau$-axis
of Fig.~\ref{fig:lambda_tau}. In particular we derive conditions for
the unstable dimensions of the trivial steady state near the Hopf
bifurcation point $\lambda=0$ in our model equation~(\ref{eq:sub}). On
the $\tau$-axis ($\lambda=0$), the characteristic
equation~(\ref{eq:char}) for $\eta=i\omega$ is reduced to
\begin{eqnarray}
\eta = i + b\left( e^{-\eta \tau}-1 \right),
\label{eq:char2}
\end{eqnarray} 
and we obtain two series of
Hopf points given by 
\begin{eqnarray}
0\leq\tau_n^A&=&2\pi n\\
0<\tau_n^B&=&\frac{2\beta+2\pi n}{1-2 b_0 \sin\beta}\quad (n=0,1,2,\dots).
\end{eqnarray} 
The corresponding Hopf frequencies are $\omega^A=1$ and $\omega^B=1-2
b_0 \sin\beta$, respectively.  Note that series A consists of all
Pyragas points, since $\tau_n^A=nT=\frac{2\pi n}{\omega^A}$. In the
series B the integers $n$ have to be chosen such that the delay
$\tau_n^B\geq 0$.  The case $b_0 \sin\beta = 1/2$, only, corresponds
to $\omega^B=0$ and does not occur for finite delays $\tau$.  

We evaluate the crossing directions of the critical Hopf eigenvalues
next, along the positive $\tau$-axis and for both series.
Abbreviating $\frac{\partial}{\partial \tau} \eta$ by $\eta_\tau$ the
crossing direction is given by $\text{sign}(\Re~\eta_\tau)$. Implicit
differentiation of (\ref{eq:char2}) with respect to $\tau$ at
$\eta=i\omega$ implies
\begin{eqnarray}
        \text{sign}(\Re~\eta_\tau)=-\text{sign}(\omega) \, \text{sign}(\sin(\omega\tau-\beta)).\label{eq:bed}
\end{eqnarray} 
We are interested specifically in the Pyragas-Hopf points of series A
(marked by red dots in Fig.1) where $\tau=\tau_n^A=2\pi n$ and
$\omega= \omega^A=1$. Indeed $\text{sign}(\Re~\eta_\tau)=
\text{sign}(\sin\beta) >0$ holds, provided we assume $0<\beta<\pi$,
\textit{i.e.}, $b_I>0$ for the feedback gain. This condition alone,
however, is not sufficient to guarantee stability of the steady state
for $\tau< 2n\pi$.  We also have to consider the crossing direction
$\text{sign}(\Re~\eta_\tau)$ along series B, $\omega^B=1-2 b_0
\sin\beta$, $\omega^B\tau_n^B= 2\beta+2\pi n$, for $0<\beta<\pi$.
Eq.~(\ref{eq:bed}) now implies $\text{sign}(\Re~\eta_\tau)=
\text{sign}((2 b_0 \sin\beta-1)\sin\beta)$.  

To compensate for the destabilization of $z = 0$ upon each crossing of
any point $\tau_n^A=2\pi n$, we must require stabilization
($\text{sign}(\Re~\eta_\tau)<0$) at each point $\tau_n^B$ of series B.
This requires $0<\beta < \arcsin\left(1/(2 b_0)\right)$ or
$\pi-\arcsin\left(1/(2 b_0)\right)<\beta < \pi$. The distance between
two successive points $\tau_n^B$ and $\tau_{n+1}^B$ is
$2\pi/\omega^B>2\pi$. Therefore, there is at most one $\tau_n^B$
between any two successive Hopf points of series A. Stabilization
requires exactly one such $\tau_n^B$, specifically:
$\tau_{k-1}^A<\tau_{k-1}^B<\tau_k^A$ for all $k=1,2,\dots$,n. This
condition is satisfied if, and only if,
\begin{equation}
        0<\beta<\beta_n^*,  \label{eq:bed1}
\end{equation}
where $0<\beta^*_n<\pi$ is the unique solution of the transcendental
equation
\begin{eqnarray}
\frac{1}{\pi}\beta^*_n+2nb_0 \sin\beta^*_n = 1. \label{eq:b_star}
\end{eqnarray} 
This holds because the condition $\tau_{k-1}^A<\tau_{k-1}^B<\tau_k^A$
first fails when $\tau_{k-1}^B=\tau_k^A$. Eq.(\ref{eq:bed1})
represents a necessary but not sufficient condition that the Pyragas
choice $\tau_P=nT$ for the delay time will stabilize the periodic
orbit.  

\begin{figure}[ht] 
  \epsfxsize=\linewidth \epsfbox{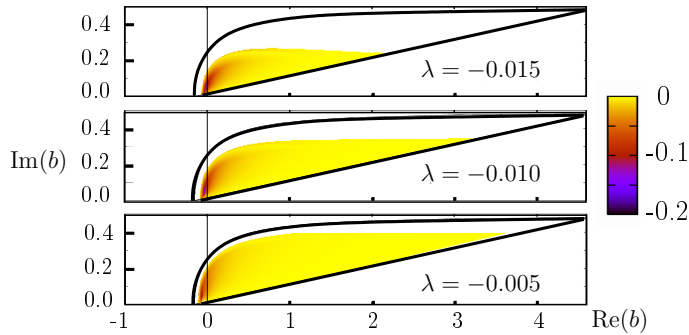}
        \caption{(Color online) Domain of control in the plane of the complex feedback 
          gain $b=b_0 e^{i\beta}$ for three different values of the
          bifurcation parameter $\lambda$. The black solid curves
          indicate the boundary of stability in the limit $\lambda
          \nearrow 0$, see (\ref{eq:bed3}), (\ref{eq:b_0_upper}).  The
          color-shading shows the magnitude of the largest (negative)
          real part of the Floquet exponents of the periodic orbit
          ($\gamma = -10$, $\tau=\frac{2\pi}{1-\gamma \lambda}$).
        }\label{fig:3}
\end{figure}

To evaluate the second condition, $\tau_H<\tau_P$ near
$(\lambda,\tau)=(0,2\pi)$, we expand the exponential in the
characteristic eq. (\ref{eq:char}) for $\omega \tau \approx 2\pi n$,
and obtain the approximate Hopf curve for small $|\lambda|$:
\begin{equation}
        \tau_H(\lambda) \approx 2\pi n- \frac{1}{b_I}(2\pi n b_R + 1)\lambda.
\label{eq:linHopf}
\end{equation}
Recalling (\ref{eq:tau_Pyragas}), the Pyragas stabilization condition
$\tau_H(\lambda)<\tau_P(\lambda)$ is therefore satisfied for
$\lambda<0$ if, and only if,
\begin{equation}
        \frac{1}{b_I}\left( b_R+\frac{1}{2\pi n} \right)<-\gamma.  \label{eq:bed2}
\end{equation} 
Eq.(\ref{eq:bed2}) defines a domain in the plane of the complex
feedback gain $b=b_R+ib_I=b_0 e^{i\beta}$ bounded from below (for
$\gamma<0<b_I$) by the straight line
\begin{equation}
        b_I=\frac{1}{-\gamma}\left(b_R+\frac{1}{2\pi n} \right).  \label{eq:bed3}
\end{equation} 
Eq.~(\ref{eq:b_star}) represents a curve $b_0(\beta)$, \textit{i.e.},
\begin{eqnarray}
        b_0 = \frac{1}{2 n \sin\beta}\left( 1-\frac{\beta}{\pi}\right), \label{eq:b_0_upper}
\end{eqnarray} 
which forms the upper boundary of a domain given by the inequality
(\ref{eq:bed1}). Thus (\ref{eq:bed3}) and (\ref{eq:b_0_upper})
describe the boundaries of the domain of control in the complex plane
of the feedback gain $b$ in the limit of small $\lambda$.
Fig.\ref{fig:3} depicts this domain of control for $n=1$,
\textit{i.e.}, a time-delay $\tau=\frac{2 \pi}{1-\gamma \lambda}$. The
lower and upper solid curves correspond to Eq.~(\ref{eq:bed3}) and
Eq.~(\ref{eq:b_0_upper}), respectively. The color code displays the
numerical result of the largest real part, wherever $<0$, of the
Floquet exponent, 
calculated from linearization of the amplitude and phase equations around 
the periodic orbit. Outside the color shaded areas the periodic
orbit is not stabilized. With increasing $|\lambda|$ the domain of
stabilization shrinks, as the deviations from the linear approximation
(\ref{eq:linHopf}) become larger. For sufficiently large $|\lambda|$
stabilization is no longer possible,
in agreement with Fig.\ref{fig:lambda_tau}(b).
Note that for real values of $b$, i.e., $\beta=0$, no stabilization occurs at all.
Hence, stabilization fails if the feedback matrix $B$ is a multiple of the
identity matrix. 

In conclusion, we have refuted a theorem which claims that a periodic
orbit with an odd number of real Floquet multipliers greater than
unity can never be stabilized by time-delayed feedback control. For
this purpose we have analysed the generic example of the normal form
of a subcritical Hopf bifurcation, which is paradigmatic for a large
class of nonlinear systems. We have worked out explicit analytical
conditions for stabilization of the periodic orbit generated by a
subcritical Hopf bifurcation in terms of the amplitude and the phase
of the feedback control gain \cite{ccdomain}. Our results underline
the crucial role of a non-vanishing phase of the control signal for
stabilization of periodic orbits violating the odd number limitation.
The feedback phase is readily accessible and can be adjusted, for
instance, in laser systems, where subcritical Hopf bifurcation
scenarios are abundant and Pyragas control can be realized via
coupling to an external Fabry-Perot resonator \cite{SCH06a}. The
importance of the feedback phase for the stabilization of steady
states in lasers \cite{SCH06a} and neural systems \cite{ROS04}, as
well as for stabilization of periodic orbits by a time-delayed
feedback control scheme using spatio-temporal filtering \cite{BAB02},
has been noted recently.  Here, we have shown that the odd number
limitation does not hold in general, which opens up fundamental
questions as well as a wide range of applications.  The result will
not only be important for practical applications in physical sciences,
technology, and life sciences, where one might often desire to
stabilize periodic orbits with an odd number of positive Floquet
exponents, but also for tracking of unstable orbits and bifurcation
analysis using time-delayed feedback control \cite{SIE06}.
  
\acknowledgements This work was supported by Deutsche
Forschungsgemeinschaft in the framework of Sfb 555.
        

\begin{thebibliography}{10}
\bibitem{ccdomain}
For the complex conjugate values of $b$, stabilization of the periodic orbit
can be shown by analogous arguments.
\end{thebibliography}

\end{document}